\documentstyle[mncite,psfig]{mn}
\newcommand{\etacar}{$\eta$ Carinae}
\def\kms{\ifmmode{~{\rm km\,s}^{-1}}\else{~km~s$^{-1}$}\fi}
\def \nii {[N~{\sc ii}]\ $\lambda$6584~\AA}

\newcommand{\ha}{H$\alpha$} 
\newcommand{\degree}{$^{o}$~}

\def\vsys{\ifmmode{V_{\rm sys}}\else{$V_{\rm sys}$}\fi}

\newcommand{\vhel}{{V_{\rm hel}}}
\title[Strings in $\eta$ Carinae]
{The origin of the strings in the outer regions of $\eta$ Carinae}
\author[M.P. Redman, J. Meaburn and A.J. Holloway]
{M.P. Redman$^{1,2}$, J. Meaburn$^1$ and A.J. Holloway$^1$\\ 
$^1$Jodrell Bank Observatory, University of Manchester, Macclesfield SK11
9DL, UK\\
$^2$Department of Physics and Astronomy, University College London, Gower
Street, London WC1E 6BT, UK}

\date{Received **insert**; in original form **insert**}
\pagerange{\pageref{firstpage}--\pageref{lastpage}}

\begin{document}
\label{firstpage}
\maketitle

\begin{abstract} 
The narrow optical filaments (`strings' or `spikes') emerging from the
Homunculus of $\eta$ Carinae are modelled as resulting from the passage
of ballistic `bullets' of material through the dense circumstellar
environment. In this explanation, the string is the decelerating flow
of ablated gas from the bullet. An archive HST image and new forbidden
line profiles of the most distinct of the strings are presented and
discussed in terms of this simple model.
\end{abstract}


\section{Introduction} 
A recent addition to the wealth of physical features associated with
the Luminous Blue Variable star \etacar\ are the remarkable high speed
filaments protruding radially from the Homunculus and visible in deep
${\rm H}~\alpha$ and \nii\ imagery (Meaburn, Wolstencroft \& Walsh
1987\nocite{meaburn.et.al87}, \pcite{meaburn.et.al96}). These features
(which are variously referred to as `spikes' - Meaburn et al.\@~1996;
`whiskers' - Morse et al.\@~1998\nocite{morse.et.al98}; or `strings' -
Weis, Duschl \& Chu~(1999); we will use strings) are observed to
extend out into the surrounding nebulosity.

Weis et al.\@~(1999) have summarised the known properties of these
strings, of which 5 are visible in HST images. They are typically
$2-5 \times 10^{17}~{\rm cm}$ long and $6 \times 10^{15}~{\rm cm}$
wide, tapering somewhat towards the tip. This length to width ratio of
between 30-100 marks them as highly collimated structures. The strings
point directly back to \etacar. They are not perfectly straight
however as some localised kinks and brightness knots are seen in some
of the strings.

The velocity structures of the strings are no less remarkable than
their optical morphologies. \scite{meaburn.et.al96} using the EMMI
spectrometer on the New Technology Telescope (La Silla) found that,
along the length of the most prominent string (String 1), the radial
velocity changes from $\sim -630~{\rm km~s^{-1}}$ to up to $\sim
-850~{\rm km~s^{-1}}$ at the discernable tip of the string (the
systemic heliocentric radial velocity of \etacar\ $\vsys = 7
\kms$). Weis et al.\@~(1999) \nocite{weis.et.al99} were subsequently 
able to show that this velocity change is in fact linear with distance.

In this paper, new line profiles from the most prominent string are
presented and compared with a previously published position-velocity
array of line profiles from the area where the string emerges from the
Homunculus. Various possible explanations are considered for the
strings and a model of the strings based upon a dense bullet of
material ploughing through the circumstellar environment of \etacar\
is developed.

\section{Observations}

The HST archive \nii\ image of String 1 in Fig. 1 is one where the HST
diffraction spikes generated by \etacar\ are conveniently displaced so
that the string is clearly visible. Note, however, that due to the
extreme radial velocities of the strings some emission can be lost due
to the narrow bandwidth of some of the HST filters (see Meaburn et
al.\@~1996 and Weis et al.\@~1999 for a full discussion and complete
set of images of the strings). The position/velocity (pv) array of
\nii\ line profiles (see Meaburn et al.\@~1987 for details) in Fig. 2
is from the east/west slit length intersecting position A 12\arcsec\
south of \etacar, where String 1 emerges from the inner circumstellar
shell that surrounds the Homunculus. The \nii\ and \ha\ line profiles
in Figs. 3 a \& b are from positions B and C respectively (see Fig. 1)
along String 1 (see Meaburn et al.\@~1996 for the technical
details). These have been extracted from the longslit pv arrays and
corrected for the emission from adjacent regions along the slit
length. The very high radial velocities of these features can cause
contamination from nearby spectral lines. The different spectral lines
displayed for the two positions were chosen to avoid this.

\begin{figure}
\psfig{file=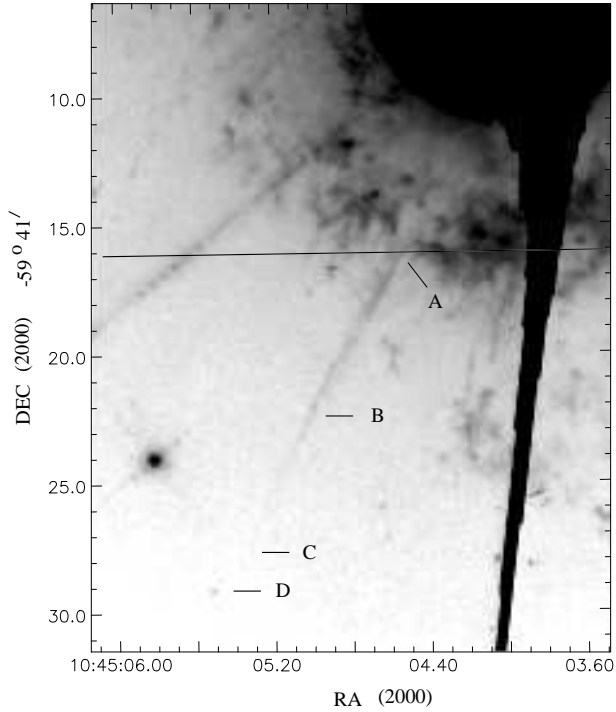,width=250pt,bbllx=0pt,bblly=90pt,bburx=500pt,bbury=650pt}
\caption{HST archive image of String 1 in the light of \nii. The slit positions 
of Meaburn et al.\@~1996 are labelled and discussed in the text. The
overexposed region at the top of the figure is the Homunculus with
$\eta$ Carinae itself lying beyond the top right corner of the image.}
\label{hst}
\end{figure}
\begin{figure}
\psfig{file=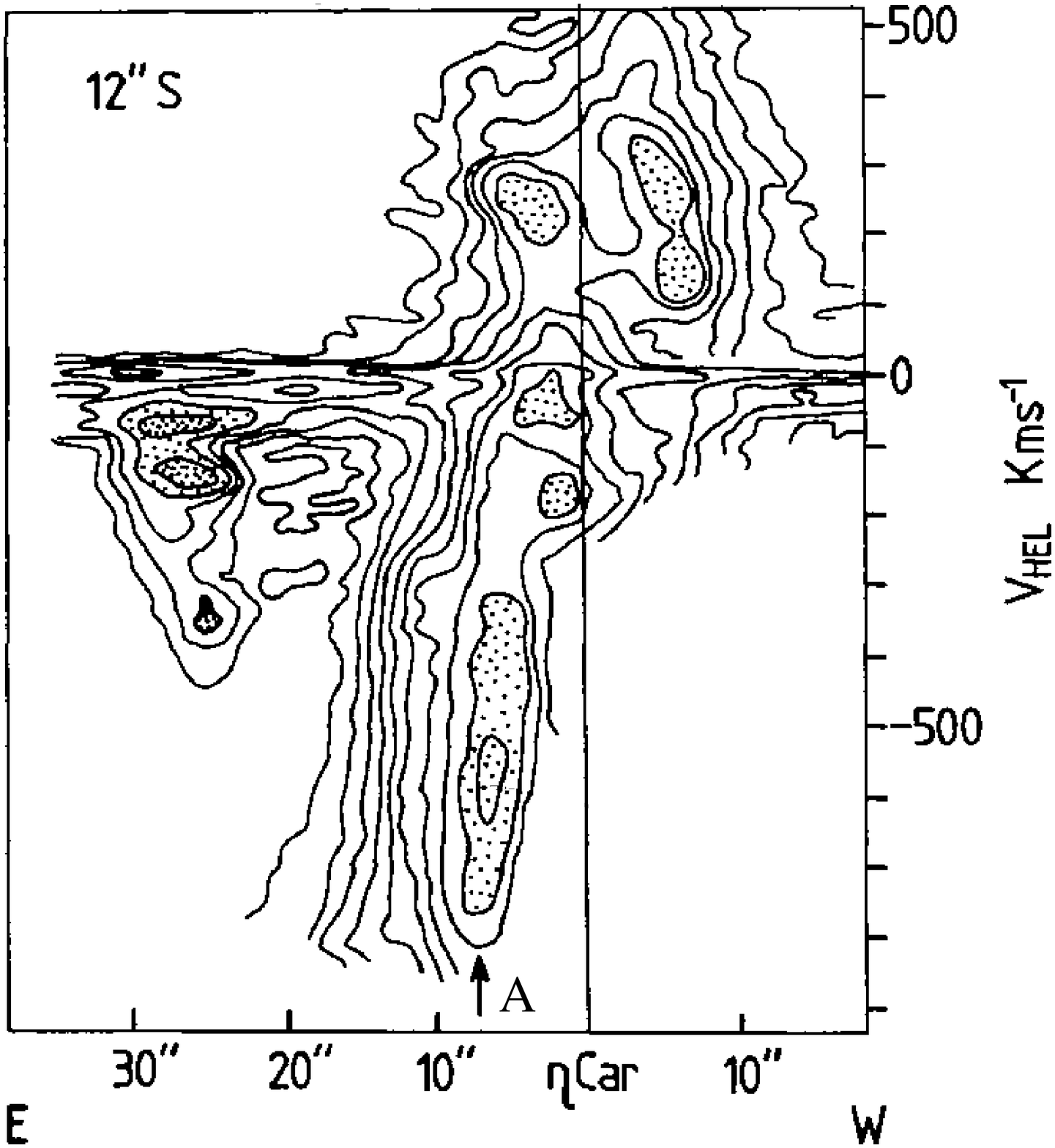,width=250pt,bbllx=0pt,bblly=0pt,bburx=761pt,bbury=725pt}
\caption{Position velocity arrays of line profiles obtained from the slit 
position that passes through point A, marked in Figure 1. The label `$\eta$ Car' 
marks the point of the slit immediately due south of $\eta$ Carinae.}
\label{fig2}
\end{figure}

It can be seen in the pv array in Fig. 2 that the base of String 1 at
A coincides closely with a remarkable spatially localised velocity
feature that extends from $\vhel= -800~{\rm km~s^{-1}}$ (at the bottom
of the figure) to \vsys. This large range of velocities over indicates
that the region at which the string emerges from the inner shell of
\etacar\ is highly disturbed.

The profiles from String 1 at positions C and B (Fig. 3 a \& b) are
easier to interpret. Profile C is clearly split by $\approx$ 48 \kms\
while profile B seems to consist of two blended components of
comparable separation. These imply that the string is expanding at 24 \kms\
perpendicularly to the length of the string. However, the centroid of
the split from position B is shifted by --600 \kms\ from \vsys\ while
that from position C is shifted by --860 \kms. Positions B and C are
at 21\arcsec\ and 27\arcsec\ from
\etacar\ respectively so a reasonably linear change of radial velocity
along the length of String 1 is indicated.  Even with such large
shifts in radial velocity the widths (FWHM) of the individual velocity
components (corrected for the 12 \kms\ instrumental resolution) are
only $\approx 40 \kms$.

\begin{figure}
\psfig{file=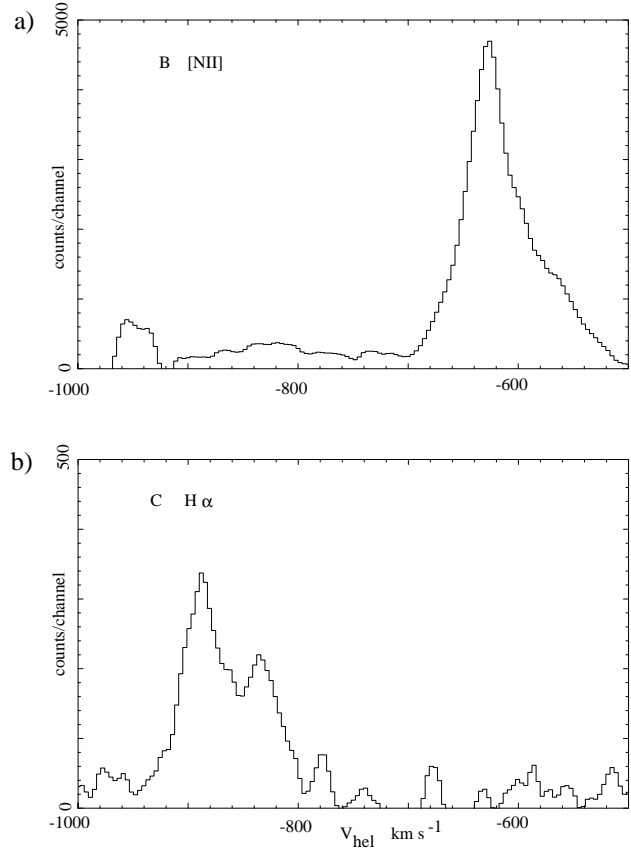,width=250pt,bbllx=0pt,bblly=0pt,bburx=541pt,bbury=724pt}
\caption{An \nii\ and an \ha\ line profile from (a) position B 
and (b) position C respectively}
\label{fig3}
\end{figure}

Note that there is an unresolved stellar-like feature (at D in Fig. 1)
and, although somewhat detached, it is in line with String 1 and
\etacar. Currie et al.\@~(2000b) have used proper motion measurements
to show that this is a bullet travelling at a higher velocity than the
rest of the string and is likely to be physically associated with the
string. Their astrometric measurements suggest that the bullets are
travelling at up to $3000~{\rm km~s^{-1}}$ i.e. around $1\%$ of the
speed of light. The angle of the string to the plane of sky is
determined by comparison of the astrometric velocity measurements with
radial velocity measurements to be 25\degree (Currie et
al.\@~2000b). This angle agrees with that estimated by Weis et
al.\@~(1999).

\begin{figure}
\psfig{file=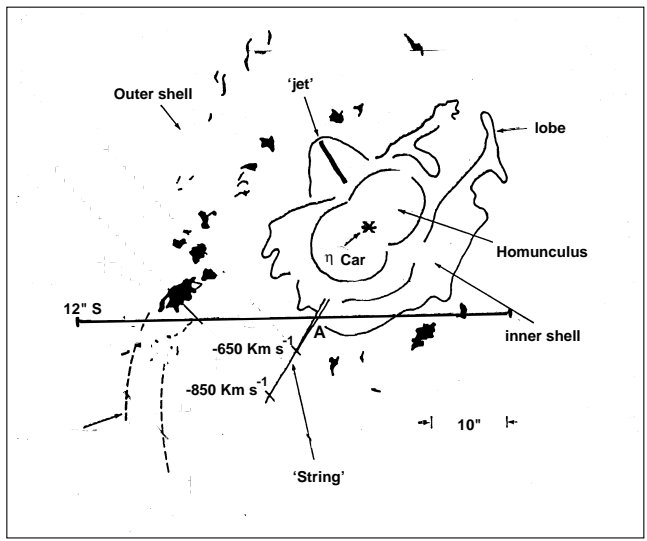,width=250pt,bbllx=0pt,bblly=0pt,bburx=553pt,bbury=553pt}
\caption{Sketch of the main features of the $\eta$ Carinae nebulosity. The slit 
position marked is orientated exactly east-west}
\label{fig4}
\end{figure}

\section{Possible explanations}

In this section, various possible explanations for the origin of the
strings are considered in turn. \scite{garcia_segura.et.al99} have
shown that, for a planetary nebula medium in which the magnetic
pressure dominates the gas pressure, narrow collimated jet like
features with a velocity that increases along the jet can be
produced. The calculations were two dimensional and they argued that
when extended to 3D, pinching instabilities could lead to localised
disturbances along the jet and possibly break up into clumps.

At first sight, this type of model, suitably adapted for the
conditions in the $\eta$ Carinae nebulosity, appears promising.  A
major problem is that a bipolar axis is required along which these
structures will form. In $\eta$ Carinae there are at least 5
strings. There are likely to be more since the observed strings happen
to lie at fortuitous positions and face-on and obscured strings will
not be detected. Some kind of precessing axis also appears unlikely
since the dynamical times of the strings indicates that they were all
ejected at the time of the great outburst of the 1840s although a
rapidly precessing and tumbling axis cannot be ruled out.

The number of strings is also a problem for a hydrodynamical jet
interpretation for similar reasons to those above. Several collimating
sources with differing orientations are required and there is also the
problem of the seemingly coincident temporal origin. Again, some kind
of source that varied rapidly within $\eta$ Carinae remains a
possibility. It is also worth noting that there is an episodic jet
emerging from the Homunculus \cite{meaburn.et.al93b} but it bears
little similarity to the strings (it is much broader and brighter and
has a single velocity). Amongst a list of suggestions, Weis et
al.\@~(1999) suggest that the strings could result from projection
effects from larger funnels but as discussed here, generating numerous
such funnels would require a jet and the only certain jet feature is
not attended by strings.

In favourable circumstances, one could envisige gaps in a
circumstellar shell allowing beams of emission to be formed as the
radiation scatters off dust in the beam. Indeed, the striking infrared
image of the Homunculus by \scite{currie.et.al00a} seems to show just
this effect: the Homunculus appears covered in short spikes of
emission. However, this is easily ruled out as a possible explanation
for the optical strings here since the beams should be straight rather
than kinked as observed.

The most likely explanation is one which involves the ejecta of the
1840s outburst (as concluded by Weis et al.\@~1999 and
\pcite{currie.et.al00b}) since the ages of the strings and their
distribution about the Homunculus is then accounted for. One of the
possibilities listed by Weis et al.\@~(1999) is that the strings could
be a train of individual bullets following the same path. This cannot
be firmly ruled out until higher resolution observations become
available but there are problems with such an interpretation. The
strings are of uniform kinematical age, so at an earlier time the
bullets must have all been located at the same position and have been
travelling in the same direction yet with a range of velocities. This
could be the result if a large fragment of ejecta breaks up but the
velocity and mass distribution of the resulting fragments must then be
fairly uniform to allow the strings to remain coherent and
recognisible. Furthermore, each fragment must retain the exact
direction of motion of the original body as it breaks up. These
constraints make it seem unlikely that the strings result from a train
of bullets.

\scite{soker01} 
has suggested that the strings are the result of ionization shadows
behind dense clumps in the Homunculus. The surrounding gas is ionized
and compresses the shadow gas which is then ionized sometime later,
becoming visible as a string. The velocity structure of the strings is
not accounted for by the model. The discovery of the high speed knots
of emission beyond the tips of the strings (see Currie et al.\@~2000a)
may cause problems for this model since this suggests that the strings
may have been somehow generated behind a high velocity clump rather
than through the more quiescent mechanism postulated by Soker (though
the model may still be applicable to the filaments in NGC 5443).

\scite{redman&meaburn01} described a model for the strings
in which the strings are generated by single bullets ploughing through
the surrounding medium. The trails of material left behind as the
bullet is disrupted are identified as the strings. We develop this
model in the next section. Weis~et~al.\@~(1999) included the
possibility that the spikes could be generated by such a mechanism in
their list of possible explanations.

\section{A model for the strings}

We suggest that the strings are simply the effects of the passage of
dense fragments of ejecta interacting with the circumstellar
environment. Knots of supernova ejecta travelling at higher speeds
than the edge of the supernova remnant have been found to lie beyond
the main remnant body
(e.\@g.\@~Braun~et~al~1987\nocite{braun.et.al87}). We suggest that a
similar process operates in $\eta$ Carinae so that the densest pieces
of ejecta will be slowed negligibly and overrun the main Homunculus
body. The visible string is the trail left behind by the passage of
the ejected fragment. In \scite{dyson.et.al93} and subsequent papers
the problem of the tail shape that results from an interaction between
a clump and a flow has been studied in detail. However,
\scite{dyson.et.al93} are concerned with quasi-steady flows that can
occur after the initial stages of the interaction of the clump and
flow have settled down and so their results are probably not
applicable here. \scite{klein.et.al94} have carried out a detailed
analytical and numerical study of the interaction between a small
clump and a strong shock and they trace the evolution of the clump as
it is hit by the shockwave, embedded in the post-shock flow and
eventually destroyed. The timescale for the destruction of the clump
is shorter than the time taken for a clump to sweep up a column
density of material from the ISM because the knots are vulnerable to
the growth of Kelvin-Helmholtz (K-H) instabilities when embedded in
the post shock flow. Defining $\chi\equiv {n_{\rm c}/n_{\rm i}}$ as
the density contrast between the clump and the ISM, with $n_{\rm c}$
and $n_{\rm i}$ being the clump and ISM densities respectively, the
destruction time, $t_{\rm d}$, is then \cite{klein.et.al94}

\begin{equation}
t_{\rm d}=\chi \frac{r_{\rm c}}{V_{\rm s}},
\end{equation}

\noindent 
where $V_{\rm s}$ is the shock speed. Applying this cloud destruction
equation to the bullets from \etacar, $V_{\rm s}\la 1000~{\rm
km~s^{-1}}$ is used as an upper limit to the shock speed (the
deprojected velocity of the highest speed string material - Section
1). The bullet radius is $r_{\rm c}\simeq 3 \times 10^{15}~{\rm cm}$
i.e. assuming the initial bullet radius to be approximately half that
of the width of the trail (see below). Since the bullet appears to
still survive, the destruction time $t_{\rm d}$ must be greater than
150 years. This gives that the density constrast between the clump and
external medium should be $\chi \ga 100$. Observationally, this
requirement seems plausible. \scite{kennicutt84} measured the density
of the Carina nebula to be $\sim 100~{\rm cm^{-3}}$ towards the centre
of the nebula. \scite{morse.et.al98} estimate that dusty clumps in the
Homunculus have densities of $10^{5}-10^{6}$. They also measure the
density of the strings to be $400~{\rm cm^{-3}}$. The strings have a
length to width ratio of 30-100 so assuming the bullet to originally
be comparable in diameter to the string, suggests an initial bullet
density of a $\ga 10^4~{\rm cm^{-3}}$. It seems reasonable then to
adopt a bullet density of $10^4~{\rm cm^{-3}}$ and an ambient density
of $10^2~{\rm cm^{-3}}$.

That the trail of material ablated from the bullet has not appreciably
expanded is easily verified because the dynamical age of the trail is
$t_{\rm dyn}\simeq 150~{\rm yr}$ (similar to that of the Homunculus)
and even if the oldest parts of the tail had been expanding at the
thermal sound speed of ionized gas ($c_{\rm i}=10~{\rm km~s^{-1}}$)
for this time they would only have swelled to
\begin{equation}
c_{\rm i}t_{\rm dyn}\la 4.5\times 10^{15}~{\rm cm},
\end{equation}
The distance to \etacar\ used is 2.3~kpc (this is an unambiguous
distance measurement - see \pcite{meaburn99}). The expansion is less
than the thickness of the strings of $6\times 10^{15}~{\rm cm}$
estimated by Weis et al.\@~(1999), who also point out that there is
little evidence that the string gets progressively thinner towards its
tip. It is unlikely therefore that the clump had an initial diameter
greater than the width of the string. Since the clump is detected in
the HST images, it seems reasonable to adopt a clump diameter of the
order of the thickness of the string.

Within the strings, localised knots of emission are seen in HST
images. They do not appear to be any wider than the rest of the
string. Weis et al.\@~(1999) marginally resolve the velocity widths of
the localised knots to be $\simeq 22~{\rm km~s^{-1}}$ (in accord with
our measurents in Section 2). If the clump is destroyed primarily by
the development of K-H instabilities, the most disruptive modes are
those with wavelengths of the order of the original clump radius. The
embedded knots could then represent eddies in the tail generated by
the K-H instability and the velocity widths would then be a measure of
their internal turbulence. The slight deviations in the otherwise
remarkably straight tails most likely result from the highly
non-linear action of the disruption process.

The displacement of the bullet and strings (see Figure 1, and Currie
et al.\@~2000b) has a straightforward explanation in terms of our
simple model: the gas stripped from the ejecta fragment may be
initially too hot to be optically visible and only once it has cooled
somewhat downstream will it be detectable. The details of the
shock-cloud interaction will be complicated but we envisage that a
slow shock driving into the dense bullet gives rise to the optically
visible \nii\ emission at the bullet surface while a fast bow shock
between the heated gas expanding from the bullet surface and the
incoming medium would heat this material to a high temperature. Some
of the gas will be heated to $\ga 10^6 {\rm K}$ as it is exposed to a
$1000 \kms$ shock but some portions of the shock surface will be
oblique to the incoming flow, giving a range of post-shock
temperatures.

The detailed formation of the string itself may be as follows. Hot
postshock gas is stripped from the fast moving clump/bullet. The
temperature of this gas depends on the very uncertain shock
strength. It is likely to be in the range $10^5-10^6~{\rm K}$. In
order to produce optical emission, this gas must cool through UV
resonance lines. The detection of O~{\sc vi} resonance lines would be
a simple test that this cooling is occuring since this UV line
indicates the presence of gas with a temperature of $\sim 10^5 {\rm
K}$. Unfortunately, the ejecta is of low O abundance so this
observation is unfeasible at present. The cooling rate of gas in the
temperature range $5\times 10^4~{\rm K} - 5\times 10^7~{\rm K}$ can be
well approximated by
\cite{kahn76}
\begin{equation}
\Lambda=1.33\times 10^{-19} T^{-1/2}~{\rm erg~cm^3~s^{-1}}.
\end{equation}
Once the lower end of this temperature range is reached, the gas will
primarily cool through optical forbidden lines. The narrow velocity
widths of $\simeq 24~{\rm km~s^{-1}}$ (Section 2) for the strings are
indicitive of a local sound speed of ionized gas at a temperature of
$\sim 10^4~{\rm K}$.

If the standard shock results apply for the post shock density and
pressure ($P_{\rm s}={3/4}\rho_0 V_{\rm s}$; $\rho_{\rm s}=4\rho_0$)
the cooling time of the shocked bullet gas is \cite{kahn98}
\begin{equation}
t_{\rm cool} = \frac{P_{\rm s}^{3/2}}{q\rho_{\rm s}^{5/2}} =\left(
\frac{3}{4}\right)^{3/2}\left( \frac{1}{32} \right) q^{-1} \rho_0^{-1}
V_{\rm s}^3, 
\end{equation}
where $\rho_0$ is the pre-shock density and $q$ is a constant (see
Kahn 1998).

Using the following illustrative values $V_{\rm s}\simeq 1000~{\rm
km~s^{-1}}$ (and this is an upper limit, as described above); $q\simeq
4\times 10^{32}~{\rm cm^6~gm^{-1}~s^{-4}}$ and $\rho_0\simeq 10^4
\times {2\times 10^{-24}}~{\rm gm~cm^{-3}}$ the cooling time is then
\begin{equation}
t_{\rm cool}\la 80~{\rm yrs}.
\end{equation}

The displacement between the string and bullet is roughly 5 percent of
the length of the string (see Fig 1). Since the string is 150 years
old, a cooling time of $\sim 10~{\rm yrs}$ is required. Note that the
shock velocity used is an estimated upper limit. The medium into which
the strings are moving is likely to be flowing away from the star,
which will reduce the velocity difference and hence shock strength
(halving the shock velocity to $500~{\rm km~s^{-1}}$ is required to
give a cooling time of $t_{\rm cool}\sim 10~{\rm yrs}$). Given the
highly uncertain shock speeds and densities this very simple model
gives a very reasonable quantative agreement with the available
observations.

Finally, the velocity structure of the strings may be accounted for as
follows. Away from the head of the string, the material in the string
will be moving at a different velocity than that of the neighbouring
medium. There will be a gradual slowing of the edges of the string at
the interface between the string gas and external medium. If this is a
steady deceleration process, then clearly the older parts of the
string will have been slowed more than those recently stripped from
the bullet and the velocity along the string will therefore change in
a linear fashion. The disturbed part of the inner shell at the base of
the string (Fig 2, Section 2) may represent the effects of the bullet
breaking through the inner regions of the nebula.

In the absence of measured values for many of the physical quantities,
the model cannot be developed much further. Given the uncertainties,
we suggest that our very simple model is a plausible explanation for
the origin and structure of the strings.

\section{Conclusions}
We have suggested that the narrow, collimated strings emerging from
the inner shell of $\eta$ Carinae are due to the passage of dense
fragments of ejecta through the circumstellar environment. We have
argued that this very simple model is capable of broadly accounting
for most of the observed features of the strings. The clump that
formed the string is visble beyond the end of the string; the
collimation of the string is due to both the high density contrast
($\ga 100$) between the clump and the circumstellar environment and
the high speed of the clump - the string has not had time to expand
appreciably; the distance between the observed clump and string is due
to the time required for the newly shocked ablated gas from the clump
to cool and become optically visible; the linear change in velocity
along the string is due simply to the gas in the string being further
decelerated by the external environment.

This model could be developed further using the latest numerical
hydrodynamic codes to model the complicated physics involved in the
destruction of the ejecta fragments. While the $\eta$ Carinae strings are
presently unusual enough to warrant special explanation, it is
entirely possible that they may turn out to be a common feature of new
explosively generated circumstellar nebulosity.

\section*{Acknowledgements}
We thank the referee for a prompt and constructive report. We also
thank Robin Williams for useful discussions and suggestions and Jeremy
Yates for a critical reading of the manuscript. MPR is supported by
PPARC.


\label{lastpage} 
\end{document}